\documentclass[10pt]{article}
\usepackage{fancyhdr}
\usepackage{amsmath}
\usepackage{amsthm}
\usepackage{amsfonts}
\usepackage{siunitx}
\usepackage{tikz}
\usepackage[plain]{algorithm}
\usepackage{algpseudocode}
\usepackage{multirow}
\usepackage{booktabs}
\usepackage{graphicx}
\usepackage{subfigure}
\usepackage{xcolor}

\usepackage{longtable}
\usepackage{bm}
\usepackage{mathtools}
\usepackage{amssymb}
\usepackage[labelfont=bf]{caption}
\usepackage{capt-of}
\usepackage{mciteplus}
\usepackage{cite}

\usepackage{mathrsfs}
\usepackage[title,titletoc,toc]{appendix}
\usepackage{xr}
\usepackage{parskip}
\usepackage{soul}
\usepackage{textcomp}
\usepackage[colaction]{multicol}
\usepackage[switch]{lineno}
\usepackage{etoolbox}
\usepackage{array}
\usepackage{tablefootnote}
\usepackage{ragged2e}
\newcolumntype{C}[1]{>{\centering\arraybackslash}p{#1}}
\usetikzlibrary{automata,positioning}
\usepackage{adjustbox}
\usepackage{makecell}
\usepackage{pdflscape}
\usepackage{tabularx}
\usepackage{setspace}
\usepackage[T1]{fontenc}
\usepackage{textgreek}
\usepackage{xr-hyper}    
\usepackage{xurl}         

\usepackage[colorlinks=true,linkcolor=black,anchorcolor=black,citecolor=black,urlcolor=blue]{hyperref}

\theoremstyle{plain}

\theoremstyle{definition}

\newcommand{\keywords}[1]{\noindent\textbf{Keywords:} #1\par\vspace{0.5em}}



\begin{document}

\title{Multiscale Topology-Regulated Nonlinear Dynamics for Protein B-Factor Prediction and Single-Cell Type Classification}

\author{Zihan Wang$^{1}$,
Mushal Zia$^2$, Xiaoqi Wei$^2$, Li Shen$^2$,  and Jian Jiang$^{1,2}$\footnote{Corresponding author. Email: jjiang@wtu.edu.cn} \\
$^{1}$Research Center of Nonlinear Science, School of Mathematics and Statistics, \\ Wuhan Textile University, Wuhan, 430200, P. R. China \\
$^{2}$Department of Mathematics, Michigan State University,\\ East Lansing, Michigan 48824, USA }
\date{\today}

\maketitle

\begin{abstract}
Protein B-factor prediction and single-cell type classification are important for characterizing structural variability and cellular heterogeneity, with applications in biological function analysis and drug discovery. However, static geometric descriptors do not directly capture how multiscale connectivity shapes responses to local perturbations. Here, we introduce multiscale graph-coupled Burgers response learning (MGCBRL), a framework integrating algebraic topology and nonlinear dynamics to construct predictive biological representations. Specifically, a filtration converts protein coordinates or cellular principal component scores into weighted graphs at ten neighborhood scales. Independent unit impulses initiate node-specific responses governed by nonlinear Burgers dynamics. Statistical summaries or sampled source-node trajectories then encode the influence of structural neighborhoods on transient signal evolution. The simulated responses serve as computational probes rather than models of biological motion. For cross-protein prediction, multiscale response statistics are combined with spectral perturbation descriptors measuring changes in Laplacian spectra after node removal. For single-cell classification, principal component scores augment these features to retain explicit expression-space information, and the resulting descriptors support downstream supervised learning. Across 104 proteins and 13 single-cell datasets, MGCBRL achieved configuration-selected mean scores of 0.888 for within-protein Pearson correlation and 0.950 for transductive cell classification balanced accuracy, with leave-one-protein-out mean correlations of 0.496–0.558. Our framework provides a shared strategy for transforming multiscale biological connectivity into informative response descriptors, supporting predictive analysis of protein structures and transcriptomic profiles.

\end{abstract}


\tableofcontents
\newpage

\section{Introduction} \label{sec:introduction}
Accurate characterization of residue-level structural variability and cellular identity is important for understanding biological organization across molecular and cellular scales. Protein B-factor prediction provides residue-specific information associated with structural rigidity, flexibility, local disorder, and collective motion, thereby supporting the analysis of protein stability, molecular recognition, allosteric regulation, and structure-based drug discovery \cite{pandey2023b,bramer2018blind}. In parallel, cell-type classification from single-cell RNA sequencing (scRNA-seq) data is essential for resolving cellular heterogeneity, identifying rare or disease-associated cell populations, and characterizing the composition and functional organization of complex tissues \cite{tan2019singlecellnet,wang2024scfed}. Although protein structures and single-cell transcriptomic profiles differ in their physical meanings and data dimensions, both can be represented as collections of entities embedded in a geometric or feature space, with relationships determined by pairwise proximity. Their corresponding prediction tasks are also defined at the node level, estimating residue-specific B-factors in protein graphs and assigning cell-type labels in cell-cell graphs. Conventional coordinate-, distance-, or expression-based representations provide essential information but describe these systems primarily as static configurations. They do not directly characterize how a node's neighborhood redistributes an imposed perturbation or how this response changes as progressively longer-range connections are introduced. These shared limitations motivate a unified representation strategy in which static biological point clouds are transformed into multiscale graphs and studied through controlled dynamical responses \cite{jiang2025}.

Crystallographic B-factors provide a demanding setting in which to examine this idea \cite{barthels2021bandeltait, sun2019bfactor}. A range of structure-based methods has been developed to describe protein fluctuations and predict B-factors. The Gaussian network model and related elastic-network approaches approximate equilibrium fluctuations through harmonic interactions defined on residue contact graphs \cite{erman2025extending, xia2015multiscale, zheng2008unification}. Flexibility-rigidity index methods replace explicit mode decomposition with distance-dependent rigidity measures and provide efficient geometric descriptors \cite{xia2014fri, nguyen2016generalized}. Molecular nonlinear dynamics \cite{xia2014mnd} and evolutionary homology \cite{cang2020eh} approaches introduce nonlinear or evolution-based information into flexibility analysis, while topology-enabled prediction from chaos (TEPC) combines data-driven learning with dynamical features \cite{jiang2025}. More recently, persistent spectral graph and persistent sheaf Laplacian methods have demonstrated that filtration-dependent spectra encode multiscale geometric and topological information relevant to protein flexibility \cite{hayes2025psl,hayes2026predicting, wang2020psg}. Diffusion-based analysis has likewise shown that local network geometry is associated with structural flexibility \cite{cheng20254d,zhang2025flexible}. These methods establish the predictive value of contact connectivity, geometry, topology, and dynamics, but they leave room for a complementary representation based on transient, node-specific responses. Harmonic models are primarily associated with linearized fluctuations near an equilibrium configuration, whereas many geometric and spectral descriptors summarize a graph without explicitly examining nonlinear signal propagation. Moreover, a single contact threshold cannot simultaneously resolve short-range packing and longer-range structural organization. An open question is whether a nonlinear response operator applied across a graph filtration can transform these different connectivity regimes into informative and compact node-level fingerprints.

Multiscale graph construction provides a natural basis for addressing this question. Starting from pairwise distances, a filtration process generates a sequence of graphs in which short-range connections appear first and progressively longer-range relationships are subsequently incorporated. In protein structures, low filtration levels emphasize local packing, whereas higher levels increasingly represent long-range and global connectivity. The graph Laplacian associated with each level defines how a signal is redistributed through the corresponding weighted neighborhood. Applying an identical unit impulse to every node in turn therefore produces responses that depend on node degree, edge weights, local connectivity, and position within the full graph. A densely connected node can rapidly distribute the imposed signal through multiple pathways, whereas a sparsely connected node may retain a more persistent self-response. Recording this process across filtration levels converts static structural differences into trajectories whose magnitude, decay, and variability can be quantified. The resulting response descriptors complement direct geometric or spectral summaries by characterizing how each node participates in transient information transport.

The same representation principle can be tested in a domain without protein-specific spatial or sequential structure. scRNA-seq measurements describe cells through thousands of expression variables and are commonly reduced to a lower-dimensional space before clustering or classification. Recent advances in single-cell annotation include foundation models and graph-based representation learning. CellFM leveraged large-scale transcriptomic pretraining to support cell-type annotation \cite{zeng2025cellfm}. Graph-based approaches increasingly exploit adaptive connectivity and biological priors. For instance, scALGSL combined active learning with graph structure optimization to address label scarcity and noisy connections \cite{du2026scalgsl}, whereas ScanNet integrated gene expression and transcriptional regulatory networks through heterogeneous graph learning \cite{long2026scannet}. These developments motivate complementary representations that characterize node-specific responses to standardized computational perturbations across multiple graph scales.

Principal component (PC) scores capture major axes of transcriptional variation, but a cell's identity also depends on its relationship to local and multiscale neighborhoods. We therefore construct cell-cell graphs in PC space and apply the graph-coupled Burgers impulse-response procedure across the resulting filtration. This extension is evaluated as a cross-domain test. Protein C$\alpha$ coordinates describe a folded molecular structure with a natural residue order, whereas transcriptomic profiles form a feature-space point cloud without an equivalent biological ordering. Success in both settings would not establish universal applicability, but it would indicate that multiscale graph responses can provide useful node representations beyond the domain in which the framework was initially formulated.

In this work, we introduce multiscale graph-coupled Burgers response learning (MGCBRL). The classical Burgers operator combines nonlinear advection with viscous dissipation, providing a compact mechanism for generating amplitude- and gradient-dependent responses \cite{burgers1948mathematical,wei2002conjugate}. The Burgers-inspired evolution is treated as a computational probe. The response generated from each node reflects its local neighborhood, broader graph connectivity, and their changes across spatial scales. We evaluate MGCBRL through three related settings. First, we perform within-protein B-factor regression on 104 proteins spanning three size ranges, comparing statistical and trajectory-based Burgers representations with established structure-based approaches and examining how filtration level, coupling strength, and regression model affect performance. Second, we conduct leave-one-protein-out (LOPO) prediction to assess whether information encoded by multiscale Burgers statistics and node-specific Laplacian spectral perturbations transfers to proteins excluded from model fitting. Third, we evaluate cell-type classification on 13 scRNA-seq datasets using a combined representation of multiscale response statistics, spectral perturbation sensitivities, and PC scores. The experiments show strong configuration-selected within-protein B-factor correlations and positive mean correlations under LOPO evaluation. For single-cell classification, the combined representation exceeds the PC comparator across all 13 datasets, with ablations supporting multiscale aggregation and PC-score augmentation under the evaluated protocol. Together, these results establish MGCBRL as a reproducible strategy for converting static protein and transcriptomic point clouds into task-relevant node descriptors, providing a generalizable approach for linking spatial organization with molecular function and improving data-driven analysis in structural biology and single-cell research.

\section{Results}\label{sec:results}

The results cover within-protein B-factor regression, cross-protein prediction under the LOPO protocol, and scRNA-seq cell-type classification.  Figure~\ref{fig:framework} outlines MGCBRL, and Figure~\ref{fig:results-overview} summarizes predictive performance across these settings. 

\subsection{Overview of the framework}\label{subsec:overview}

\begin{figure}[H]
\centering
\includegraphics[width=0.99\textwidth]{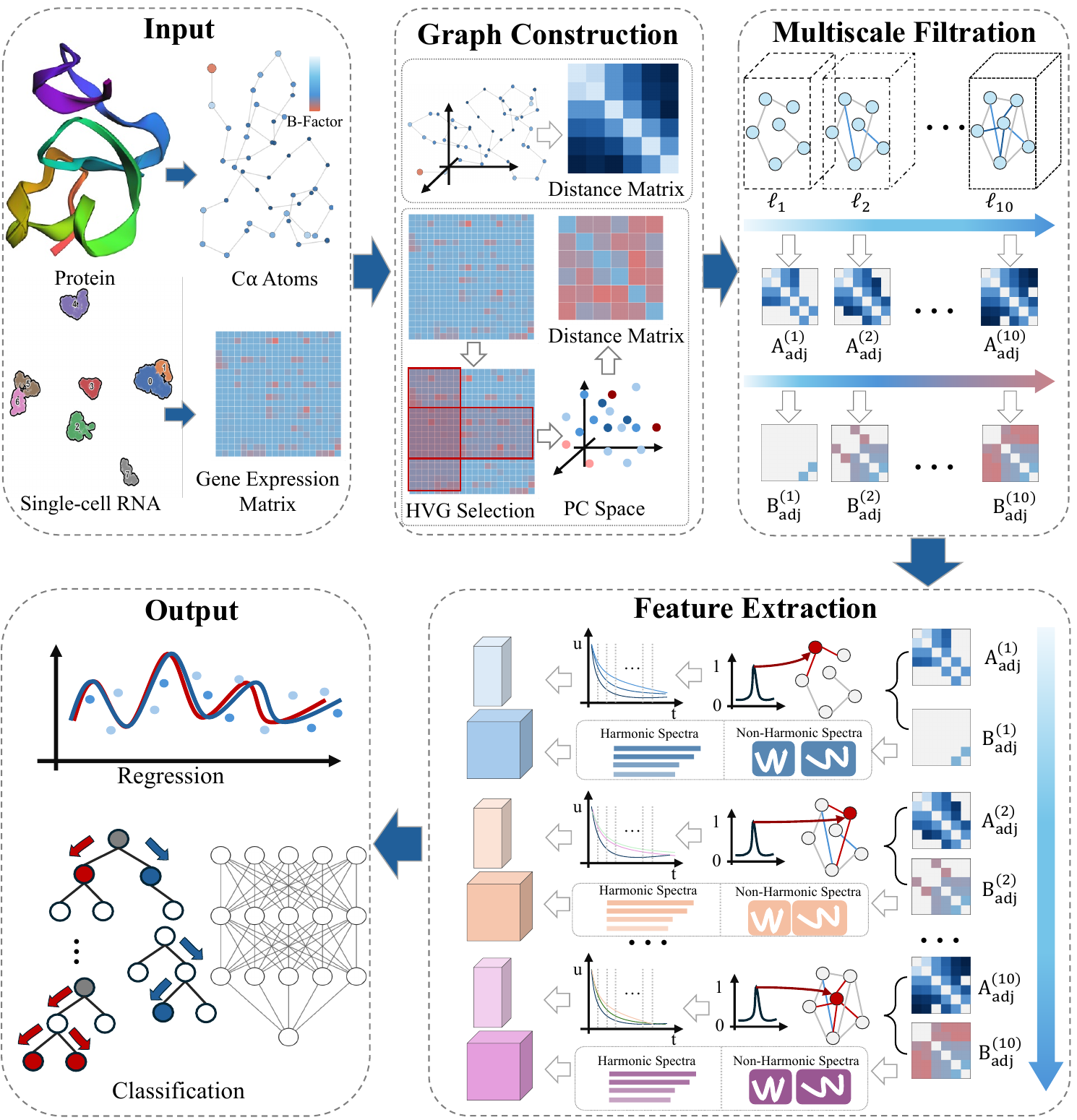}
\caption{Workflow of multiscale graph-coupled Burgers response learning for protein B-factor regression and single-cell classification. Pairwise Euclidean distances are computed from protein Cα coordinates or cell-specific scores in a 30-dimensional principal-component (PC) space obtained after expression preprocessing and highly variable gene (HVG) selection. Percentile-based filtration process produces ten weighted graphs, represented by adjacency matrices $A_{\mathrm{adj}}^{(\ell)}$ for proteins and $B_{\mathrm{adj}}^{(\ell)}$ for cells, where $\ell=1,\ldots,10$. Each node serves in turn as the source of a unit impulse in a separate graph-coupled Burgers simulation, yielding a source-node response trajectory. Spectral perturbation features are computed separately as absolute changes in Laplacian zero-eigenvalue multiplicity (harmonic component) and positive-eigenvalue summaries (non-harmonic component) following node removal. Within-protein regression uses statistical response summaries or sampled trajectories. Cross-protein regression combines multiscale response statistics with spectral perturbation features, whereas cell-type classification uses the corresponding cell-level features together with the 30 PC scores.}
\label{fig:framework}
\end{figure}

\begin{figure}[!htbp]
\centering
\includegraphics[width=0.99\textwidth]{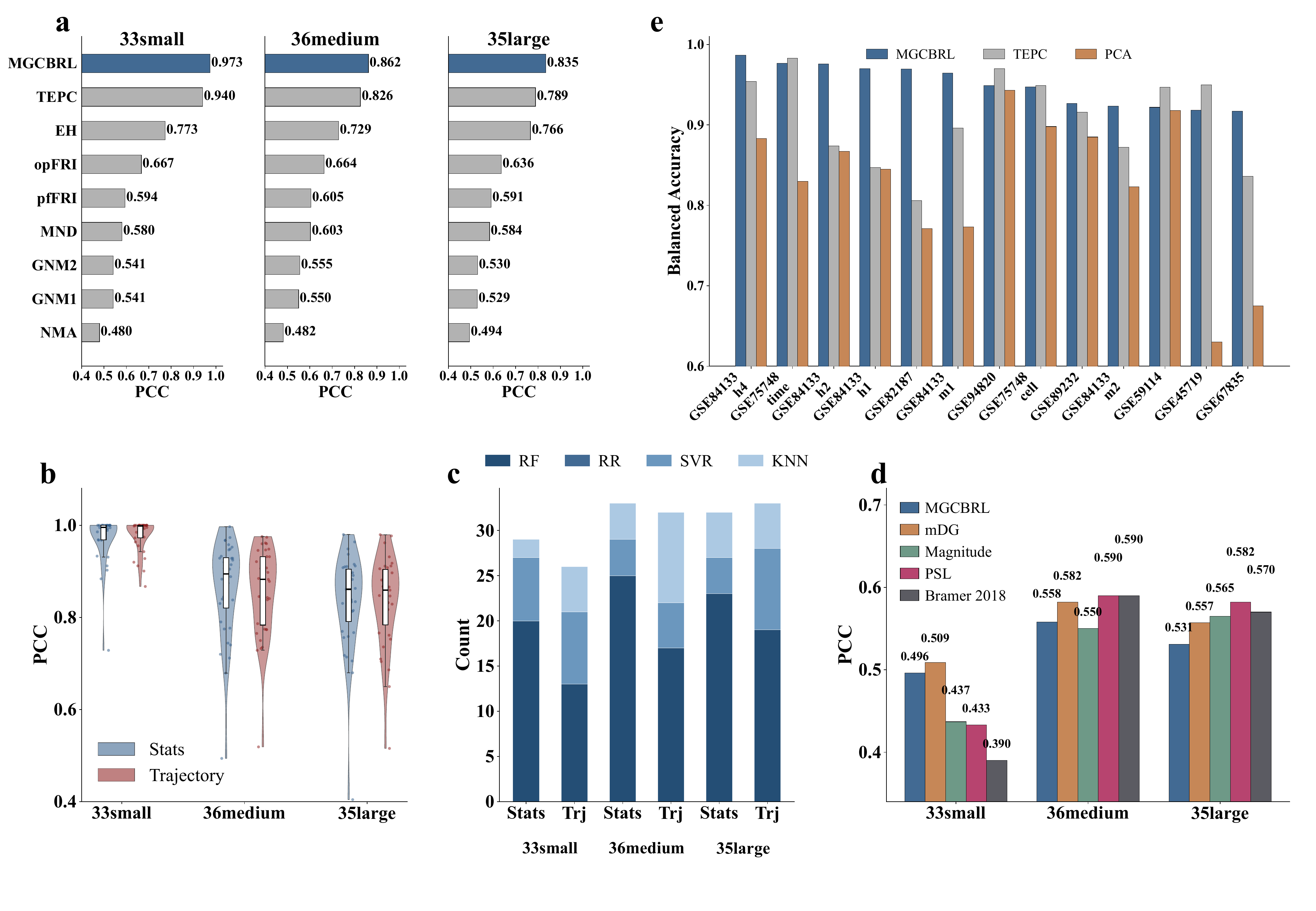}
\caption{Predictive performance for protein B-factor regression and single cell type classification. (a) Comparison of the mean PCC obtained by our method (MGCBRL) under the 80\%/20\% train-test split with values for eight baseline methods across the 33small, 36medium, and 35large datasets. (b) Per-protein PCC distributions for statistical dynamical features (Stats) and trajectory dynamical features (Trj), shown as violin plots with box plots and individual points. (c) Selection frequencies of random forest (RF), ridge regression (RR), support vector regression (SVR), and $k$-nearest neighbors (KNN) for each representation and dataset. (d) Mean PCC obtained by MGCBRL using gradient-boosted decision trees (GBDT) under the leave-one-protein-out protocol, compared with multiscale differential geometry (mDG), Magnitude, the persistent sheaf Laplacian (PSL), and Bramer2018  \cite{bramer2018blind}. (e) Balanced accuracy (BA) of MGCBRL, TEPC, and principal component analysis (PCA) for cell-type classification across 13 scRNA-seq datasets. }
\label{fig:results-overview}
\end{figure}

MGCBRL combines multiscale graph construction with Burgers impulse-response analysis for protein B-factor regression and scRNA-seq cell-type classification, which is illustrated in Figure~\ref{fig:framework}. Proteins are represented by their C$\alpha$ coordinates, whereas cells are represented by their scores along the first 30 principal components of the preprocessed gene expression matrix after highly variable gene (HVG) selection. Pairwise Euclidean distances define a distance matrix in each coordinate spaces. Filtration at 5th, 15th, …, and 95th percentiles of the off-diagonal distance distribution yields ten nested graphs. Consequently, the resulting weighted adjacency matrices are denoted by $A_{\mathrm{adj}}^{(\ell)}$ for proteins and $B_{\mathrm{adj}}^{(\ell)}$ for cells, where $\ell=1,\ldots,10$. At each filtration level $\ell$, each node is treated as a nonlinear oscillator and embedded a nonlinear dynamical system, such as the Burgers dynamics \cite{burgers1948mathematical}, serving in turn as the source of a unit impulse in an independent graph-coupled Burgers simulation. After filtration, a family of dynamical systems is constructed using topological Laplacians \cite{wang2020psg}, which encode the interactions into multiscale distance matrices for dynamical systems. Subsequently, the source-node response trajectories are then converted into descriptors that form the dynamical component of the feature sets used for regression or classification.


Persistent spectral theory extracts topological and geometric information from the filtered graphs. Its harmonic and non-harmonic spectra describe topological invariants and geometric characteristics, respectively, extending the information captured by persistent homology in topological data analysis \cite{suTopo2025,weeRev2025}. These spectra provide global graph descriptors. Node-specific spectral perturbation features are obtained by topological differentiation, which measures changes in the descriptors after node removal. Concatenating these features across filtration levels yields a multiscale representation for downstream learning. Dynamical features characterize the generated responses, whereas spectral perturbation features summarize how each node contributes to graph structure.

MGCBRL was evaluated in three settings with different feature combinations and validation protocols. Within-protein B-factor regression compared two representation methods of dynamical features, six-statistic response summaries and 100-point sampled trajectories at individual filtration levels under the nominal 80\%/20\% train-test split (See Sections 4.2.5-4.2.6). Cross-protein prediction combined dynamical feature with spectral perturbation feature, specifically, 60 multiscale Burgers statistics with 60 residue-specific Laplacian spectral perturbation features, yielding a 120-dimensional feature (See Table~S4 of the Supporting Information). LOPO evaluation excluded the held-out protein from both coupling-strength selection and model fitting (See Section 4.3). For scRNA-seq classification, 30 cell-specific PC scores were appended to the corresponding 120-dimensional representation. The resulting 150-dimensional vectors were evaluated using repeated stratified cross-validation (CV), as detailed in Section 4.4 and Table~S17 of the Supporting Information.

We want to emphasize that the coupled Burgers equation used in MGCBRL serves as a controllable nonlinear feature extractor, without assuming that protein atoms obey Burgers dynamics or exhibit shock-like motion. Each node carries a dimensionless local perturbation amplitude, whose evolution converts static structural connectivity into a dynamic response feature. The coupling term in Burgers equation aggregates neighboring state differences, providing a computational analogy to message passing in graph neural networks. Across filtration levels, the responses characterize how different structural neighborhoods influence signal retention and decay, providing features whose associations with B-factors are learned from data. Additionally, for a specified graph and fixed dynamical parameters, the Burgers response trajectories of dynamical feature were computed without using crystallographic B-factors or cell-type annotations. These labels were used for supervised model fitting, configuration selection, and performance evaluation under the protocols described below. The simulated trajectories describe the evolution of an algorithmic signal rather than atomic motion or transcriptional dynamics. Their association with residue-level B-factors or cell identity was assessed through the corresponding prediction tasks.

\subsection{Within-protein B-factor prediction}\label{subsec:within-protein}

Within-protein B-factor prediction was evaluated on 104 proteins from the 33small, 36medium, and 35large datasets using six-dimensional statistical dynamical features (Stats) and 100-dimensional trajectory dynamical features (Trj) at individual filtration levels. The reported values represent configuration-selected within-protein performance.

Figure~\ref{fig:results-overview}a shows the prediction results under the 80\%/20\% train-test split and compares the mean PCC values of our method (MGCBRL) with those of eight baseline methods, including topology-enabled predictions from chaos (TEPC)~\cite{jiang2025}, evolutionary homology (EH)~\cite{cang2020eh}, the optimal flexibility-rigidity index (opFRI) and parameter-free flexibility-rigidity index (pfFRI)~\cite{xia2014fri}, molecular nonlinear dynamics (MND) ~\cite{xia2014mnd}, two Gaussian network model variants (GNM1 and GNM2)~\cite{xia2014fri}, and normal mode analysis (NMA)~\cite{park2013gnm}. These methods span chaotic-learning, topology- and dynamics-based, flexibility-rigidity, and harmonic elastic-network approaches. We find that using statistical dynamical features, our method MGCBRL outperforms all other eight methods and achieves the largest mean PCCs for all three datasets. Specifically, MGCBRL obtains mean PCC values of 0.973 ± 0.054, 0.862 ± 0.105, and 0.835 ± 0.107 (mean ± standard deviation (SD)) across proteins in the 33small, 36medium, and 35large datasets, which are improved up to 102.7\%, 78.8\%, and 69.0\% compared with those of 0.480, 0.482, and 0.494 in \cite{park2013gnm}, respectively. Additionally, the corresponding 95\% confidence intervals (CIs) for the mean PCCs were 0.954–0.993, 0.827–0.897, and 0.798–0.872, and the mean PCCs across all 104 proteins was 0.888. 

Performance nevertheless varied across the three datasets. The lower means and larger standard deviations in 36medium and 35large indicate less consistent predictions across their constituent proteins. Greater heterogeneity in structural neighborhoods may contribute to this variation, although evaluation-subset size and configuration selection also affect the comparison. A potential strength of the present approach is its ability to select a filtration scale and coupling strength for each protein, allowing the statistical dynamical features to reflect different connectivity regimes. This adaptability may help accommodate protein-specific structural environments. The details of mean simulation and regression runtimes of prediction are reported in Table~S22 of the Supporting Information.

Figure~\ref{fig:results-overview}b compares the distributions of within-protein PCC values obtained using statistical and trajectory dynamical features. Both representations produced correlations concentrated near 1.0 in 33small, whereas 36medium and 35large showed broader distributions with longer lower tails, indicating greater variation among proteins. The two representations nevertheless achieved similar dataset-level means. Trajectory dynamical features yielded a slightly higher mean PCC in 33small, at 0.978 compared with 0.973 for statistical dynamical features, while statistical dynamical features performed slightly better in 36medium, at 0.862 compared with 0.857. Both yielded a mean PCC of 0.835 in 35large at the reported precision. The details of these comparisons can be found in Table S15 of the Supporting Information. These small differences indicate that neither representation achieved higher mean performance across the three datasets.

Trajectory dynamical features exhibited lower cross-protein standard deviations in all three datasets, suggesting somewhat more consistent scores under the evaluated protocol. Retaining temporal profiles may help distinguish responses with similar summary statistics, although the close mean PCC values suggest that much of the predictive information is already retained by the six statistical descriptors. Statistical dynamical features therefore offer a compact representation with similar average performance to the 100-dimensional trajectories. The broader distributions in the medium and large datasets may reflect more heterogeneous structural environments, while the near-ceiling scores in 33small also require consideration of small evaluation subsets and configuration selection. Overall, Figure~\ref{fig:results-overview}b supports statistical summaries as an efficient representation and identifies modestly lower cross-protein variability as a potential benefit of retaining the full sampled response profile. Detailed results for statistical dynamical features and trajectory dynamical features are provided in Tables~S9 and S10 for 33small, Tables~S13 and S14 for 36medium, and Tables~S11 and S12 for 35large, all of which are given in the Supporting Information.

In Figure~\ref{fig:results-overview}c, RF was the most frequently selected regressor for both feature representations in all three datasets, although its predominance was greater for statistical dynamical features. With statistical dynamical features, RF was selected for 20 of 33 small proteins, 25 of 36 medium proteins, and 23 of 35 large proteins, accounting for 68 of 104 proteins in total. For trajectory dynamical features, the corresponding counts fell to 13, 17, and 19, totaling 49 proteins. Meanwhile, selections of SVR and KNN increased from 15 to 22 and from 11 to 20, respectively, whereas RR remained comparatively infrequent. Thus, retaining the sampled temporal profiles changed the distribution of preferred regressors without displacing RF as the most common choice. Complete model-selection counts for both statistical dynamical features and trajectory dynamical features are reported in Table~S8 of the Supporting Information.

\subsection{Cross-protein B-factor prediction}\label{subsec:blind-results}

Cross-protein B-factor prediction was evaluated separately within each dataset using LOPO validation. The results were averaged over five LOPO cycles with different random seeds. In Figure~\ref{fig:results-overview}d, MGCBRL with GBDT yielded mean PCC values of 0.496 ± 0.318, 0.558 ± 0.179, and 0.531 ± 0.122 for 33small, 36medium, and 35large, respectively. The corresponding 95\% CIs for the mean PCCs were 0.383–0.608, 0.498–0.619, and 0.489–0.573. Its strongest relative performance occurred in 33small, where it ranked second behind mDG \cite{feng2025mdg} and exceeded Magnitude \cite{bi2025topomag}, PSL \cite{hayes2025psl}, and Bramer2018 \cite{bramer2018blind} by 0.059, 0.063, and 0.106. In 36medium, MGCBRL slightly exceeded Magnitude but remained below mDG, PSL, and Bramer2018. For 35large, its mean PCC was lower than those of all four comparators, with differences ranging from 0.026 to 0.051. These rankings indicate that the favorable within-protein results did not translate into a consistent advantage when predicting proteins excluded from model fitting. Additionally, dataset-level PCC and root mean square error (RMSE) results for GBDT, RF, RR, and KNN are reported in Table~S7, and the fixed GBDT hyperparameters are provided in Table S6 of the Supporting Information.


Figure~\ref{fig:results-overview}d revealed a different performance pattern from the within-protein analysis. The lower cross-protein PCC values may reflect the difficulty of transferring structure–B-factor relationships between proteins with different structural environments and experimental conditions. Within-protein models learn from residues in the same structure, whereas LOPO prediction must accommodate differences between training and test proteins. The comparison also involves different evaluation procedures. Within-protein configurations were selected using the reported evaluation PCC, which can inflate performance, particularly for small evaluation subsets. LOPO excluded test-protein labels from both coupling-strength selection and model fitting. Differences in feature representations and regressors further contribute to the comparison, preventing attribution of the performance gap solely to cross-protein transfer.

\subsection{scRNA-seq cell-type classification}\label{subsec:scrna}

MGCBRL was evaluated for cell-type classification on 13 scRNA-seq datasets containing 300–1,937 cells. The final 150-dimensional representation combined 60 Burgers response statistics, 60 spectral perturbation features, and 30 cell-specific PC scores. Classification followed the transductive within-dataset protocol described in Section 4.4, with all cells included in unsupervised preprocessing and feature construction. The four-model selection procedure (ML4) selected the coupling strength and classifier using repeated stratified five-fold CV. Because the same CV results were used for configuration selection and performance summarization, the reported BA values represent configuration-selected performance.

In Figure~\ref{fig:results-overview}e, MGCBRL achieved the highest mean BA across the 13 scRNA-seq datasets, with 0.950 ± 0.026 compared with 0.908 ± 0.057 for TEPC \cite{jiang2025} and 0.826 ± 0.093 for PCA. The corresponding 95\% CIs for the mean BA were 0.934–0.965, 0.874–0.942, and 0.770–0.882, respectively. We find that MGCBRL exceeded PCA on every dataset and maintained BA above 0.900 throughout, indicating high configuration-selected performance across the evaluated datasets. Its smaller cross-dataset standard deviation also indicates less variation in classification performance. The improvement over PCA was particularly pronounced for GSE45719 and GSE67835, where BA increased from 0.630 to 0.918 and from 0.675 to 0.917, respectively. In contrast, the difference was only 0.004 for GSE59114, suggesting that the additional representation offered limited benefit where the PCA comparator already performed well.

The comparison with TEPC was more dataset dependent. MGCBRL achieved higher BA on eight datasets and lower BA on five, with the largest increase observed for GSE82187, from 0.806 to 0.969. Thus, its higher overall mean reflects substantial improvements on several datasets rather than a uniform advantage over TEPC. A plausible explanation is that combining multiscale response and spectral descriptors with PC scores captures neighborhood information that is not explicit in individual cell coordinates, while retaining major axes of expression variation. The usefulness of this additional information may vary with dataset-specific cell organization and class separability.  Figure~\ref{fig:results-overview}e supports the combined representation as a promising approach to within-dataset cell-type classification, although the transductive, configuration-selected evaluation does not establish performance on independent datasets. Per-dataset results and the experimental configuration are reported in Tables~S16 and S17 of the Supporting Information, respectively.

The four configurations in Figure 3a distinguish the effects of feature augmentation and classification strategy. Mean BA increased from 0.864 for 120-dimensional fixed RF to 0.918 for 150-dimensional long short-term memory (LSTM), 0.930 for 150-dimensional fixed RF, and 0.950 for 150-dimensional ML4. Using the same input information, fixed RF exceeded LSTM on 11 of the 13 datasets. The details of per-dataset BA values for the four configurations can be found in Table~S19 of the Supporting Information. This suggested that recurrent processing of the filtration sequence was not generally advantageous under the tested settings. Direct access to the concatenated descriptors may allow RF to exploit useful combinations across scales without compressing them into a final recurrent state. Nevertheless, LSTM performed better on GSE89232 and GSE59114, indicating that the relative suitability of the two classifiers varied among datasets. The LSTM architecture and training settings are provided in Table~S18 of the Supporting Information.

\begin{figure}[H]
\centering
\includegraphics[width=0.98\textwidth]{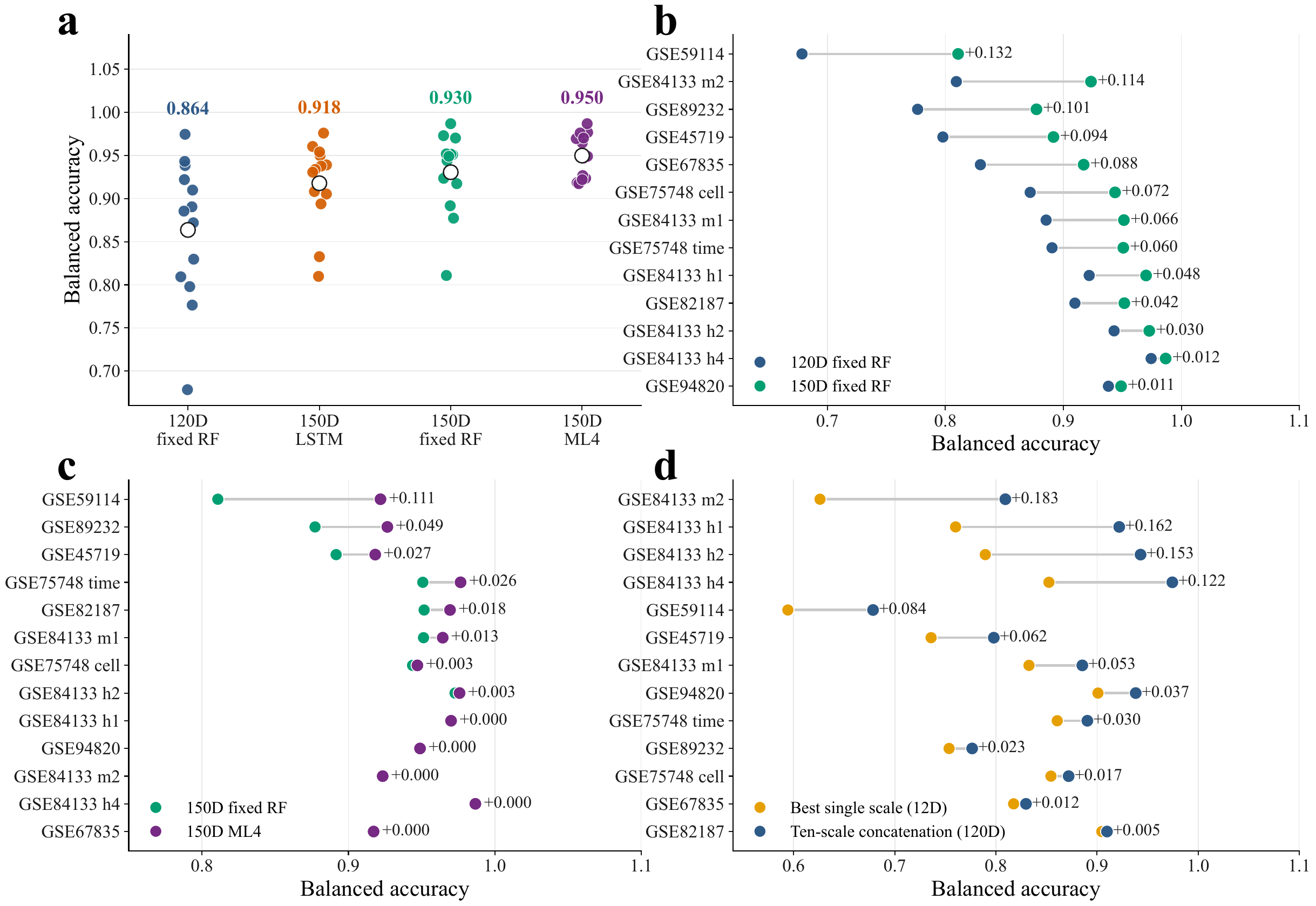}
\caption{Ablation analysis of the scRNA-seq feature assembly and classification strategy. (a) Paired balanced-accuracy distributions across 13 datasets for the 120-dimensional fixed-RF baseline, 150-dimensional LSTM, 150-dimensional fixed RF, and 150-dimensional ML4 configurations, and open circles denote means. (b) Per-dataset BA difference after concatenating the 30 cell-specific PC scores with the 120-dimensional multiscale representation under a fixed RF classifier. (c) Per-dataset BA difference between ML4 and fixed RF using the same 150-dimensional representation. (d) Comparison of the strongest single-scale 12-dimensional representation with the ten-scale concatenated 120-dimensional representation under fixed RF. Numerical labels in panels (b)--(d) denote signed BA differences.}
\label{fig:scrna-ablation}
\end{figure}

In Figure~\ref{fig:scrna-ablation}b, appending the 30 PC scores improved fixed-RF performance on every dataset, with a mean BA increase of approximately 0.067. The largest increase occurred for GSE59114, at approximately 0.13, whereas GSE94820 and GSE84133h4 improved by only about 0.01. These differences suggest that explicit expression-space coordinates provide information not fully retained by the response and spectral summaries, with greater benefit in some datasets than others. The gains across all 13 datasets support retaining PC scores alongside the graph-derived descriptors.


With the 150-dimensional representation held fixed, ML4 produced higher BA than fixed RF on eight datasets and matched it on five, which is given in Figure~\ref{fig:scrna-ablation}c. The mean increase was approximately 0.019, with the largest increment of 0.111 observed for GSE59114. Model selection therefore offered a smaller average increment than PC-score augmentation, but could substantially improve individual datasets. Logistic regression was selected for seven datasets, RF for five, and KNN for one, which could be found in Table~S20 of the Supporting Information. It indicated that the combined representation did not uniformly favor a single classifier. Because RF belonged to the candidate pool and selection reused the reported CV scores, these increments characterize configuration selection rather than independently verified generalization gains.

In Figure~\ref{fig:scrna-ablation}d, multiscale concatenation improved performance on all 13 datasets. Combining all ten filtration levels increased mean BA from 0.791 for the best single-scale representation to 0.864, with improvements on all 13 datasets. The increase ranged from approximately 0.005 for GSE82187 to 0.183 for GSE84133m2. Even the best individual scale therefore failed to retain all the predictive information available across the filtration. Different neighborhood scales may capture distinct aspects of cell organization, making their joint representation useful when no single threshold adequately describes class structure. This comparison supports multiscale feature assembly, although it does not separate the contributions of response statistics and spectral perturbation features.


\section{Discussion}\label{sec:discussion}

\subsection{Response characteristics of the graph-coupled Burgers operator}\label{subsec:response-characteristics}

The graph-coupled Burgers operator transforms a protein structural graph into residue-specific impulse-response descriptors. It serves as a nonlinear computational probe, with the node state representing an algorithmic response signal rather than atomic displacement, velocity, or thermal fluctuation. Nonlinear convection and viscous diffusion act along the sequence-indexed lattice, while graph Laplacian coupling redistributes the signal through spatial contacts. These components jointly determine how a localized perturbation evolves and how its source-node trajectory encodes the surrounding structural environment.

\begin{figure}[htbp]
\centering
\includegraphics[width=0.98\textwidth]{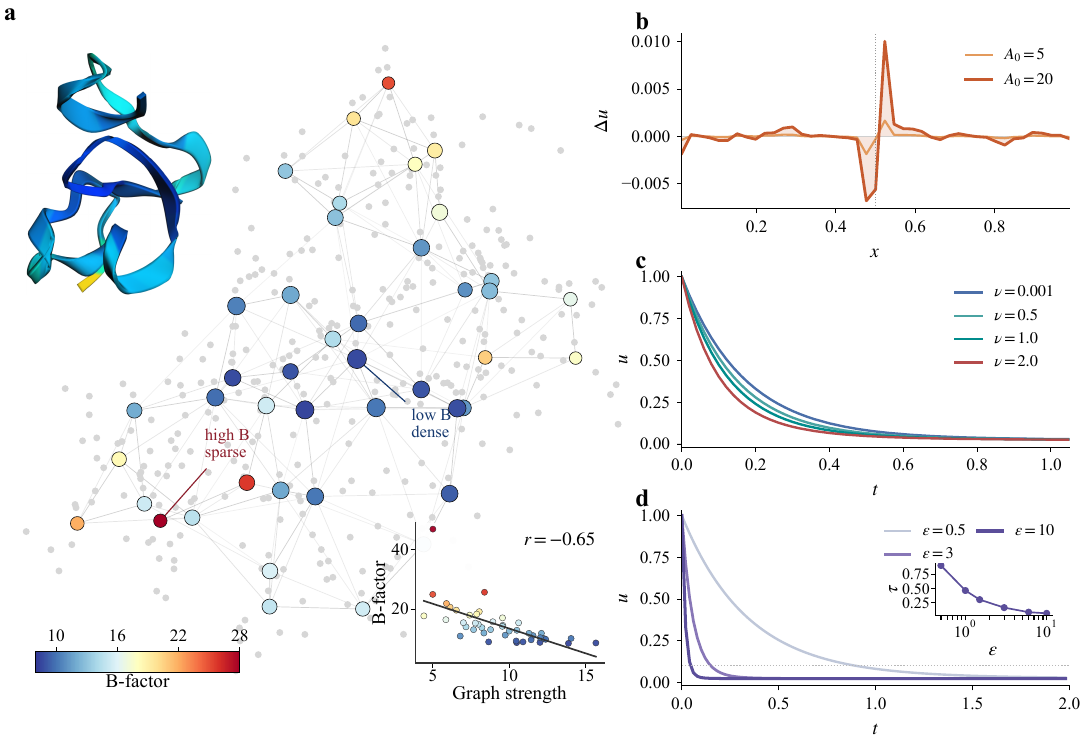}
\caption{Response characteristics of the graph-coupled Burgers operator.
(a) B-factor distribution on a two-dimensional projection of 1Q9B. Gray circles represent all atoms; colored circles represent C$\alpha$ atoms, colored by B-factor (blue, low; red, high) and sized by graph degree. Gray lines indicate the strongest 30\% C$\alpha$--C$\alpha$ edges. Insets show representative high-B-factor (sparse) and low-B-factor (dense) neighborhoods, and graph strength versus B-factor (Pearson coefficient coefficient $r \approx -0.65$). The upper-left panel shows the three-dimensional structure of 1Q9B.  (b) Response residuals for initial amplitudes $A_0 = 5, 20$, showing nonlinear amplitude dependence. (c) Source-node response $u_i(t)$ with increasing viscosity $\nu$. (d) Source-node response $u_i(t)$ with increasing coupling strength $\varepsilon$. The inset shows decay time $\tau$ versus $\varepsilon$.}
\label{fig:mechanism}
\end{figure}

In Figure~\ref{fig:mechanism}a, local weighted connectivity was inversely associated with crystallographic B-factors in the representative protein 1Q9B. Graph strength was defined as the sum of edge weights in the corresponding row of the weighted adjacency matrix in Eq.(\ref{eq6}), whose diagonal entries are zero. Because the weights are nonnegative, this quantity equals the weighted node degree and reflects both the number of retained contacts and their distance-dependent weights. Residues with greater graph strength generally had lower B-factors, yielding a Pearson correlation of approximately $-0.65$. The highlighted residues illustrate this contrast: the high-B-factor residue lies in a relatively sparse neighborhood, whereas the low-B-factor residue has denser connections.

This pattern suggests that more numerous or closer contacts may impose stronger local structural constraints, whereas weaker connectivity may permit greater positional variability. The scatter nevertheless indicates that graph strength alone does not fully account for residue-level B-factor differences. This motivates examining dynamical responses that depend on both local connectivity and signal redistribution through the surrounding network. Figure~\ref{fig:mechanism}a therefore provides a structural rationale for the response-based representation, and the association in this protein could provide a possible relationship between connectivity, response decay, and flexibility.

In Figure~\ref{fig:mechanism}b, increasing the initial impulse amplitude altered the response profile beyond proportional rescaling. This effect was assessed using the amplitude-normalized residual $\Delta u=u(A_0)/A_0-u(A_0=1)$, which would vanish for a linear response under otherwise identical conditions. For 1Q9B at filtration level L06, with $\nu=0.001$, $\varepsilon=0.5$, and $t=0.5$, the maximum absolute residual reached 2.3\% and 12.3\% of the reference profile peak for $A_0=5$ and $A_0=20$, respectively. The larger impulse therefore produced a more pronounced departure from linear scaling. A negative trough and a positive lobe appeared on opposite sides of the impulse location, indicating a spatial redistribution of the normalized signal rather than a uniform change in magnitude.

This amplitude dependence agrees with the nonlinear convection term, whose contribution depends on both the local state and its spatial gradient. Increasing the input amplitude changes its relative influence compared with the linear viscous and graph-coupling terms, allowing the operator to reshape the response rather than simply amplify it. Such behavior illustrates how the Burgers formulation can generate descriptors beyond those obtained through proportional linear diffusion. The amplitude variations here characterize the operator, whereas the prediction experiments used unit impulses. These observations establish nonlinear response behavior under the illustrated conditions.

Viscosity controlled the persistence of the source-node response in Figure~\ref{fig:mechanism}c. Increasing $\nu$ from 0.001 to 0.5, 1.0, and 2.0 produced progressively faster decay from the same initial amplitude. The curves separated most clearly during the early and intermediate stages before approaching similarly low response levels. Thus, viscosity primarily altered the transient profile over the illustrated interval, with smaller values retaining the localized signal for longer.


Stronger graph coupling markedly shortened the source-node response in Figure~\ref{fig:mechanism}d. At \(\varepsilon=0.5\), the imposed perturbation persisted for longer, whereas increasing \(\varepsilon\) to 3 and 10 produced a steeper initial decline and an earlier approach to low response levels. The inset confirms a decreasing decay time \(\tau\) across the tested coupling strengths, with smaller absolute reductions toward the upper end of the range. This pattern indicates that coupling strength provides direct control over the temporal persistence of the generated descriptors.

The observed acceleration follows from the graph Laplacian term, which redistributes the signal according to weighted differences between connected node states. Increasing \(\varepsilon\) strengthens this exchange without changing the underlying adjacency matrix, allowing the initially localized perturbation to spread more rapidly through spatial contacts, including those between residues distant in sequence. Source-node decay therefore reflects signal redistribution rather than direct measurement of physical energy loss. By changing the balance between graph-mediated exchange and lattice convection and diffusion, coupling strength adjusts how structural connectivity influences the response fingerprint. Figure~\ref{fig:mechanism}d establishes this parameter sensitivity for the illustrated configuration, but faster decay alone does not imply more informative features or better prediction. The coupling strength is consequently selected for the prediction task rather than interpreted as a physical measure of protein rigidity.

Figure~\ref{fig:mechanism} characterizes how the response dynamical features are generated. In the prediction experiments, the initial impulse amplitude and viscosity were fixed, while coupling strength was selected using predictive performance metrics. The results support the predictive utility of the evaluated configurations but do not quantify the independent contribution of each parameter. In particular, faster response decay does not necessarily imply greater predictive accuracy. A systematic assessment of how these parameters affect downstream prediction remains an unresolved aspect of the framework and a limitation of the present study.

\subsection{Interpretation of within-protein B-factor prediction}\label{subsec:interpretation}


Crystallographic B-factors reflect positional variability arising from atomic motion, static disorder, and crystallographic or modeling conditions~\cite{sun2019bfactor}. Predicting these composite observables does not necessarily require a dynamical model that reproduces their physical origins. Here, the Burgers operator transforms structural connectivity into transient response descriptors that can be associated with residue-level B-factor variation. The negative association between graph strength and B-factors in the example shown in Figure 4a provides a structural basis for this interpretation. Connectivity influences the generated responses and may also reflect constraints associated with positional variability. The resulting fingerprints can therefore carry predictive information even though their evolution occurs in an auxiliary computational system. Their relevance lies in how they encode structural environments, without implying that individual response trajectories represent actual protein motions.


The comparison between statistical and trajectory dynamical features further clarifies the value of this representation. Six statistical descriptors achieved mean PCCs close to those obtained with 100 trajectory samples across all three datasets in Figure~\ref{fig:results-overview}b. This suggests that much of the information useful under the evaluated protocol can be retained in compact summaries of the responses, while additional temporal detail does not consistently improve prediction. Such compression reduces the input dimensionality for regression, which may be useful when training data are limited within individual proteins. Variation in the selected filtration levels and coupling strengths also suggests that different proteins favor different ways of encoding their structural neighborhoods. Filtration determines which contacts enter the graph, whereas coupling strength controls their influence on the response dynamics. These parameters provide complementary flexibility in constructing descriptors, although their selected values should not be interpreted as physical interaction scales or material properties.


The strength of the reported associations should nevertheless be considered alongside the evaluation procedure. Some small proteins contributed only two evaluation residues, for which PCC is necessarily either +1 or -1 when both variables are nonconstant. In addition, filtration level, coupling strength, and regressor were selected using the same evaluation PCC subsequently reported, potentially inflating the apparent performance. These factors complicate attributing the higher correlations in smaller proteins solely to structural simplicity. Comparisons with published baseline values are also limited by differences in preprocessing and evaluation protocols. The present findings support compact transient-response descriptors as a useful representation for within-protein B-factor prediction, and their performance on residues excluded from all configuration-selection steps remains to be established.

\subsection{Cross-protein generalization}\label{subsec:cross-generalization}


Cross-protein prediction examines whether structural descriptors retain useful relationships with B-factors beyond the proteins used for training. The positive mean PCCs under LOPO evaluation indicate that the combined response–spectral representation captures some associations that extend to held-out proteins. A possible basis for this transfer is that distinct proteins can contain local connectivity patterns reflecting similar structural constraints, even when their global architectures differ. Learning from these recurring environments may therefore support prediction without access to the target protein’s B-factors. However, the variability across held-out proteins suggests that these shared relationships coexist with protein-specific effects, making transfer more successful for some structures than others.


The representation provides a practical basis for learning across proteins of different sizes. Each residue is described by the same 120-dimensional feature vector, combining response statistical dynamical features and spectral perturbation features across ten filtration levels. This common format allows residues from different proteins to enter a shared regression model without requiring residue-to-residue correspondence. The response statistics summarize how an imposed perturbation evolves, whereas the spectral features characterize changes in the graph spectrum following node removal. Their combination may describe aspects of structural organization that neither feature family captures alone, while multiple filtration levels provide access to neighborhoods of different extent. These properties motivate the feature design, although the present experiments do not isolate the contribution of each component to cross-protein performance.

A remaining challenge is that a common feature format does not guarantee a common relationship between descriptors and B-factors. Differences in structural organization and crystallographic conditions may alter this relationship across proteins. Such variation offers a plausible explanation for uneven transfer, but the reduction relative to within-protein PCCs also reflects differences in feature construction, regression models, and configuration selection. It cannot therefore be interpreted as an isolated measure of the difficulty of cross-protein prediction. The comparator rankings in Figure~\ref{fig:results-overview}d further show that the present representation does not outperform the comparators across all three datasets. The evidence supports predictive utility for held-out proteins within the evaluated collections, while transfer across broader structural and experimental diversity remains an open question.

\subsection{Cross-domain extension to single-cell transcriptomics}\label{subsec:cross-domain}

The scRNA-seq application extends the representation strategy from protein structures to cellular expression profiles. Protein graphs encode spatial relationships among residues, whereas cell graphs encode similarities in reduced expression space. In both settings, distance-based filtration defines multiscale neighborhoods from which node-specific response descriptors are constructed. The transcriptomic representation additionally incorporates spectral perturbation features and cell-specific PC scores, combining graph-derived descriptors with coordinates in reduced expression space. This extension demonstrates the applicability of a shared feature-construction strategy across two data modalities, rather than the transfer of a trained protein model to cell-type classification. Differences in graph construction, response settings, and feature assembly also make the selected coupling strengths specific to each application.


The ablations suggest that effective cell representations benefit from retaining both neighborhood information and explicit expression-space coordinates. Multiscale concatenation improved classification over the best individual filtration level, indicating that a single neighborhood threshold did not capture all the information useful for distinguishing cell types. PC-score augmentation provided an additional benefit across all datasets. Because the graph-derived features originate from the same reduced expression space, this improvement suggests that their summaries do not preserve every useful aspect of a cell’s position within that space. Combining PC scores with multiscale descriptors may therefore retain information about individual cells alongside information about their surrounding neighborhoods. The evidence supports this combined design, although the separate contributions of response statistics and spectral perturbation features remain unresolved.


\section{Methods}\label{sec:methods}

\subsection{Persistent homology and persistent spectral graph}
As a recently developed branch of algebraic topology, persistent homology (PH) provides a multiscale representation of topological features and has been successfully applied in biology \cite{xiaPersist2014}, chemistry \cite{xiaMulti2015}, and other fields \cite{weeRev2025}. Its filtration procedure constructs a family of topological spaces, enabling the system to be examined at multiple scales. However, the homotopic shape evolution of a system during filtration is not captured by PH. Persistent spectral graphs (PSGs), also known as persistent Laplacians, were subsequently developed to address this limitation \cite{wang2020psg}. Their harmonic spectra preserve the topological invariants captured by PH, whereas their non-harmonic spectra additionally characterize homotopic shape evolution during filtration. PSGs represent a special case within the broader family of persistent topological Laplacians. This family includes the persistent Hodge Laplacian on Riemannian manifolds and formulations on other topological spaces, including cellular sheaves, hyperdigraphs, and path complexes \cite{wei2025persistent}. As in PH, filtration generates successive geometric shapes and associated topological spaces on which PSGs are defined. Throughout this process, changes in the dimensions of the PSGs null spaces track the persistence of topological invariants, while the nonzero eigenvalues and corresponding eigenfunctions describe the evolution of the data’s geometric shape.

The integration of graph Laplacians with multiscale filtration led to the creation of PSGs. Spectral analysis of the $q$-combinatorial Laplacian matrix provides topological and geometric information, including the connectivity and robustness of simple graphs. PSGs generate a sequence of simplicial complexes induced by varying a filtration parameter. A filtration of an oriented simplicial complex $K$ consists of a nested sequence of subcomplexes $\{K_t\}_{t=0}^m$:
\begin{equation}
    \varnothing=K_0\subseteq K_1\subseteq K_2
\subseteq\cdots\subseteq K_m=K.
\end{equation}
For each complex $K_t$, let $C_q^t:=C_q(K_t)$ denote its $q$-chain group. The corresponding boundary operator $\partial_q^t : C_q(K_t) \to C_{q-1}(K_t)$ maps $q$-chains to $(q-1)$-chains. For $0<q\leq\dim(K_t)$, its action on an oriented $q$-simplex is given by $\partial_q^t(\sigma_q) = \sum_{i=0}^q (-1)^i \sigma_{q-1}^i$ for $\sigma_q\in K_t$, where $\sigma_q=[v_0,v_1,\ldots,v_q]$, and
$\sigma_{q-1}^{i}=[v_0,\ldots,\widehat{v_i},\ldots,v_q]$ is the oriented $(q-1)$-simplex obtained by removing vertex $v_i$. For $q<0$, $C_q(K_t)=\{\varnothing\}$, and $\partial_q^t$ is the zero map. The adjoint of the boundary operator acts in the reverse direction, $(\partial_q^t)^*:C_{q-1}(K_t)\longrightarrow C_q(K_t)$. To define the persistent boundary operator, consider the subset of $C_q^{t+p}$ consisting of chains whose boundaries lie in $C_{q-1}^{t}$, $\mathbb{C}_q^{t+p} := \left\{ \sigma \in C_q^{t+p} \mid \partial_q^{t+p}(\sigma) \in C_{q-1}^t \right\}$. On this subset, the $p$-persistent $q$-boundary operator and its adjoint are defined as $\delta_q^{t+p}:
\mathbb{C}q^{t+p}\longrightarrow C_{q-1}^{t}$ and $
(\delta_q^{t+p})^*:
C_{q-1}^{t}\longrightarrow\mathbb{C}_q^{t+p}$, respectively. The resulting $p$-persistent $q$-combinatorial Laplacian is:
\begin{equation}
   \Delta_q^{t+p} = \breve{\partial}_{q+1}^{t+p} \left( \breve{\partial}_{q+1}^{t+p} \right)^* + \left( \partial_q^t \right)^* \partial_q^t, 
\end{equation}
with the matrix representation:
\begin{equation}
    \mathcal{L}_q^{t+p} = B_{q+1}^{t+p} \left( B_{q+1}^{t+p} \right)^T + \left( B_q^t \right)^T B_q^t.
\end{equation}
The matrices $\mathcal{B}_{q+1}^{t+p}$ and $\mathcal{B}_q^t$ represent the boundary operators $\delta_{q+1}^{t+p}$ and $\partial_q^t$, respectively. The number of rows in $\mathcal{B}_{q+1}^{t+p}$ equals the number of oriented $q$-simplices in $K_t$, while its number of columns equals the dimension of $\mathbb{C}_{q+1}^{t+p}$. With respect to orthonormal bases, the transposes of these matrices represent the corresponding adjoint operators.

The Laplacian spectrum encodes topological and geometric information about the filtered complex. We denote the eigenvalues of $\mathcal{L}_q^{t+p}$ by $\left\{
(\lambda_1)_q^{t+p},
(\lambda_2)_q^{t+p},
\ldots,
(\lambda_N)_q^{t+p}
\right\},$
where $N$ is the number of rows of $\mathcal{L}_q^{t+p}$. The multiplicity of the zero eigenvalue gives the $p$-persistent $q$th Betti number,
\begin{equation}
\begin{aligned}
\beta_q^{t+p} &= \dim\left( \mathcal{L}_q^{t+p} \right) - \operatorname{rank}\left( \mathcal{L}_q^{t+p} \right) = \operatorname{nullity}\left( \mathcal{L}_q^{t+p} \right) &= \text{number of zero eigenvalues of } \mathcal{L}_q^{t+p}.
\end{aligned}
\end{equation}
Thus, $\beta_q^{t+p}$ counts the independent $q$-dimensional homology classes present in $K_t$ that persist in $K_{t+p}$. For the three-dimensional structural setting considered here, the relevant orders are $q=0,1,2$, corresponding to vertices, edges, and faces. The associated persistent Betti numbers describe the persistence of connected components, loops or tunnels, and cavities or voids, respectively. Harmonic persistent spectra track topological changes, whereas non-harmonic persistent spectra provide information about geometric evolution. By incorporating both components, this spectral framework extends the information available from persistent homology, which characterizes topological invariants.

\subsection{Protein representation and within-protein modeling}\label{subsec:bfactor-methods}
\subsubsection{Protein datasets and preprocessing}
Protein structures and their associated crystallographic B-factors were obtained from the Protein Data Bank (PDB). The analysis included three non-overlapping datasets comprising 104 proteins: 33small contained 33 proteins with 4 to 65 C$\alpha$ atoms per protein (mean 28.5), 36medium contained 36 proteins with 36 to 113 C$\alpha$ atoms (mean 90.2), and 35large contained 35 proteins with 64 to 238 C$\alpha$ atoms (mean 157.9). Dataset characteristics and processing parameters are summarized in Table~S21 of the Supporting Information.

For each protein, the three-dimensional coordinates and isotropic B-factor of each included Cα atom were extracted from the structure file. Each C$\alpha$ atom represented a residue-level graph node, and its B-factor served as the regression target. The within-protein models used response-derived descriptors without additional sequence-based or structural feature vectors. The spectral features used for cross-protein prediction are described separately in Section 4.3.

\subsubsection{Multiscale protein graph construction}

For a protein containing $N$ C$\alpha$ atoms, the pairwise Euclidean distance matrix $D\in\mathbb{R}^{N\times N}$ was computed from their three-dimensional coordinates. Ten filtration thresholds $r_\ell$ with $\ell=1,\ldots,10$ were defined by the 5th, 15th, …, and 95th percentiles of the off-diagonal distance distribution. Increasing the threshold progressively incorporated more distant residue pairs. At each level, the filtered distance matrix was defined as
\begin{equation}
    D_{ij}^{(\ell)}=
\begin{cases}
D_{ij}, & D_{ij}\le r_\ell,\\
0, & \text{otherwise}.
\end{cases}
\end{equation}

Retained residue pairs were assigned exponential distance weights. The kernel width $\sigma_\ell$ was the median of the positive off-diagonal entries of $D^{(\ell)}$ and was determined separately for each protein and filtration level. The weighted adjacency matrix was
\begin{equation}
(A_{\mathrm{adj}}^{(\ell)})_{ij}=
    \begin{cases}
 \exp\!\left(-D_{ij}^{(\ell)}/\sigma_\ell\right),
& i\ne j,\ D_{ij}^{(\ell)}>0,\\
0, & \text{otherwise}.
\end{cases}
\label{eq6}
\end{equation}
The diagonal degree matrix $D_{\mathrm{deg}}^{(\ell)}$ contained the weighted node degrees, with diagonal entries $\left(D_{\mathrm{deg}}^{(\ell)}\right)_{ii}=\sum_{j\ne i}\left(A_{\mathrm{adj}}^{(\ell)}\right)_{ij}$. The graph Laplacian was $L^{(\ell)}=D_{\mathrm{deg}}^{(\ell)}-A_{\mathrm{adj}}^{(\ell)}$, and the equivalent coupling generator was $G^{(\ell)}=-L^{(\ell)}$. Thus, $A_{\mathrm{adj}}^{(\ell)}$ had a zero diagonal, whereas $G^{(\ell)}$ had zero row sums and non-positive diagonal entries.

\subsubsection{Graph-coupled Burgers dynamics and numerical integration}
The classical one-dimensional viscous Burgers equation provides a minimal nonlinear model combining convection and diffusion \cite{burgers1948mathematical,Burgers1974,wei2002conjugate}: 
\begin{equation}
    \frac{\partial u(x,t)}{\partial t}
= -u(x,t)\frac{\partial u(x,t)}{\partial x}
+
\nu\frac{\partial^2 u(x,t)}{\partial x^2}.
\end{equation}
Here, $x$ is the one-dimensional spatial coordinate, $t$ denotes time, $u(x,t)$ is a scalar field conventionally interpreted as velocity in the fluid-mechanical setting, and $\nu>0 $ is a constant kinematic viscosity or diffusion coefficient. The term $\partial u/\partial t $ describes temporal evolution, $u\partial u/\partial x$ represents nonlinear convection with a transport speed determined by the local field value, and $\nu\partial^2u/\partial x^2$ smooths spatial variations through viscous diffusion. The competition between nonlinear steepening and diffusive smoothing produces amplitude-dependent transient responses. These properties motivate its use here as a computational feature extractor, with the field reinterpreted as a dimensionless perturbation amplitude on an auxiliary lattice and augmented by coupling through the structural graph.

In this work, the graph-coupled Burgers operator was used as a nonlinear feature extractor for protein structural graphs. Each C$_\alpha$ node carried a scalar state $u_i(t)$, defined as a dimensionless local perturbation amplitude relative to the unit input specified in Section 4.2.4. The state and its evolution describe an imposed computational signal, without assigning physical units of atomic displacement, velocity, or energy. The modeling assumption is that responses generated by a common dynamical rule can encode differences in structural neighborhoods that are informative for B-factor prediction.

Two distinct representations of the protein were used in the numerical simulation. The three-dimensional Cα coordinates determined the weighted contact graphs and their Laplacians $L^{(\ell)}$, as described in Section 4.2.2. An auxiliary one-dimensional lattice indexed the same nodes in residue order, with uniform computational spacing $\Delta x=1$. This spacing defines the discretization of the response operator and does not assume equal physical distances between C$\alpha$ atoms. Three-dimensional proximity enters through the graph coupling, whereas the lattice provides the neighboring indices required for the convection and viscous terms.

Applying second-order central differences to the spatial derivatives gave the semi-discrete system:
\begin{equation}
\frac{\mathrm{d}u_i}{\mathrm{d}t}
=
-u_i\frac{u_{i+1}-u_{i-1}}{2\Delta x}
+\nu\frac{u_{i+1}-2u_i+u_{i-1}}{(\Delta x)^2}
-\varepsilon\sum_{j=1}^{N}L_{ij}^{(\ell)}u_j,
\qquad i=1,\ldots,N.
\label{eq:protein_burgers_semidiscrete}
\end{equation}
Here, $u_i(t)$ denotes the computational response at node $i$, $\nu=0.001$ is the viscosity parameter and controls lattice smoothing, and $\varepsilon $ controls graph coupling. Periodic boundary conditions were imposed through $u_0=u_N$ and $u_{N+1}=u_1$. The resulting system of ordinary differential equations was integrated using the classical fourth-order Runge–Kutta (RK4) scheme with a fixed time step $\Delta t=0.001$. The graph term aggregates weighted differences between neighboring node states, allowing spatial contacts to transmit the signal between residues that may be distant in sequence. Varying the filtration level changes the filtered adjacency matrix and its Laplacian, generating a family of dynamical systems through which structural neighborhoods are probed at different scales.

The Burgers formulation introduces amplitude-dependent transport through the nonlinear convection term, while viscous diffusion smooths lattice variations and graph coupling redistributes the response through spatial contacts. This combination provides a compact, explicitly specified transformation of structural connectivity into temporal descriptors. Unlike linear diffusion, its responses need not scale proportionally with the imposed amplitude. This property motivates its use for feature construction, without assuming that the generated trajectories reproduce protein motion or establish the presence of shock waves in the modeled structures.

Periodic closure provides the same finite-difference stencil at every lattice site without separate endpoint rules. It is an algorithmic boundary convention and does not imply a physical contact or covalent bond between the terminal residues. It leaves the coordinate-derived distance matrix unchanged but introduces wraparound interactions in the lattice terms. Consequently, the representation depends on both the structural graph and the prescribed lattice convention, and independence from boundary treatment is not assumed.

For each filtration level and coupling strength, all $N$ node states were integrated jointly. Each node was then used in turn as the source of a separate impulse experiment, with its own response trajectory retained for feature extraction. Initial conditions, integration-horizon calibration, and response descriptors are detailed in Sections 4.2.4 and 4.2.5.

\subsubsection{Impulse-response simulations and horizon calibration}

Node-specific responses were generated by applying a unit impulse to one source node at a time. For a simulation initiated at node $i$, the initial state was:
\begin{equation}
u_j(0) = \delta_{ij} = 
    \begin{cases}
1, & j=i,\\
0, & j\ne i,
\end{cases}
\end{equation}
where $\delta_{ij}$ is the Kronecker delta. Each source node was simulated independently, and its own response trajectory was used for feature extraction.

Before the full parameter scan, a protein-specific integration horizon was calibrated at filtration level $\ell=5$ and reference coupling strength $\varepsilon_{\mathrm{ref}}=0.1$. Each node was used as the impulse source, and the first RK4 step at which its self-response fell below 0.1 was recorded as $n_i^{\mathrm{decay}}$. The number of integration steps was determined by:
\begin{equation}
   n_{\mathrm{steps}} = \min \left(
50{,}000, \max_{1\le i\le N}n_i^{\mathrm{decay}}
\right).
\end{equation}
The 50,000-step cap was used if any source response failed to reach the threshold within that limit. The same $n_{\mathrm{steps}}$ was then used for every source node, filtration level, and coupling strength within the protein. The corresponding simulation interval was $T=n_{\mathrm{steps}}\Delta t$.

The coupling strength was scanned over $\varepsilon\in\{0.1,0.2,\ldots,10.0\}$. For each combination of filtration level and coupling strength, $N$ separate impulse-response simulations were performed. The source-node trajectory was recorded over the complete integration interval for subsequent feature extraction.

\subsubsection{Dynamical feature extraction}

Two dynamical feature representations were evaluated separately: statistical summaries and sampled trajectories. (1) Statistical dynamical features consisted of the maximum, minimum, mean, variance, median, and standard deviation of the full source-node trajectory. These six summaries yielded a six-dimensional feature vector for each residue at each combination of filtration level and coupling strength. (2) Trajectory dynamical features retained the temporal profile by sampling the source-node response at 100 equally spaced points over the complete simulation interval. The sampled values formed a 100-dimensional feature vector for the same residue and parameter combination.

Within-protein regression compared the statistical and trajectory representations of dynamical features at individual filtration levels. Multiscale concatenation of the statistical dynamical features for cross-protein prediction is described in Section 4.3.

\subsubsection{Within-protein regression and evaluation}

The candidate regressors were RF, KNN, SVR, and RR. Within-protein prediction used a primary nominal 80\%/20\% train–test spliting and a supplementary adaptive protocol with CV or a holdout split according to protein size.  Model hyperparameters are summarized in Table~S21 of the Supporting Information.

(1) For each protein, graph construction and response generation used the complete set of C$\alpha$ coordinates. Residues were then partitioned into training and evaluation subsets under a nominal 80\%/20\% split. If a 20\% allocation produced fewer than two evaluation residues, two residues were assigned to evaluation and the remainder to training. A fixed random seed of 42 was used, and the same partition was applied across candidate configurations and feature representations for that protein.

PCC between predicted and observed B-factors was the primary performance measure. Candidate configurations differed in filtration level, coupling strength, and regression model. For each protein and feature representation, the configuration with the highest evaluation PCC was retained. The resulting values are configuration-selected within-protein scores and the selected configurations and per-protein results are reported in Tables~S9--S14 of the Supporting Information.

(2) The supplementary adaptive protocol used 10-fold CV for proteins with $N\ge25$, 5-fold CV for $10\le N<25$, and holdout evaluation for (N<10). For the CV cases, out-of-fold (OOF) PCC was calculated from the pooled OOF predictions for each CV-eligible protein, whereas mean-fold PCC was the arithmetic mean of its fold-specific PCC values. The holdout fallback used two evaluation residues, and these results were reported separately from the OOF summaries. Individual results and dataset-level summaries are provided in Section S1 and Tables~S1--S3 of the Supporting Information.

\subsection{Multiscale feature construction and LOPO evaluation}\label{subsec:blind-methods}

Cross-protein B-factor regression was evaluated separately within each protein dataset using LOPO validation. In each fold, one protein $h$ was reserved for testing, and the remaining proteins were used for coupling-strength selection and regressor fitting. The B-factor values of protein $h$ were used only for final evaluation. Each residue was represented by a 120-dimensional vector comprising 60 multiscale Burgers response statistics and 60 Laplacian spectral perturbation features, as summarized in Table~S4 of the Supporting Information.

The 60-dimensional Burgers dynamics feature vector was constructed from impulse responses at the ten Rips filtration levels. At each level, six summary statistics (maximum, minimum, mean, variance, median, and standard deviation) were computed from the temporal decay trajectory of the impulse-initiating atom. In each LOPO fold, one protein $h$ was held out before any label-informed configuration selection. At each filtration level $\ell$, the coupling strength was selected by maximizing the arithmetic mean of the per-protein PCC values over the remaining training proteins:
\begin{equation}
\varepsilon_{h,\ell}^{*}
=
\underset{\varepsilon\in\{0.1,0.2,\ldots,10.0\}}
{\operatorname{arg\,max}}
\left[
\frac{1}{|\mathcal{D}\setminus\{h\}|}
\sum_{p\in\mathcal{D}\setminus\{h\}}
\operatorname{PCC}_{p,\ell}(\varepsilon)
\right],
\label{eq:coupling_selection}
\end{equation}
where $\mathcal{D}$ denotes the proteins in the corresponding dataset. The resulting fold-specific schedule $\{\varepsilon_{h,\ell}^{*}\}_{\ell=1}^{10}$ was kept fixed during feature construction and GBDT model fitting for the fold in which protein $h$ was held out. Neither the B-factor values nor the PCC results of the held-out protein contributed to coupling-strength selection or model fitting. Concatenating the six statistics across the ten filtration levels yielded a 60-dimensional dynamics feature vector per atom.

To construct the 60-dimensional Laplacian spectral perturbation feature vector, spectral quantities were extracted at each of the ten filtration levels from the standard graph Laplacian $\mathcal{L}^{(\ell)}=D_{\mathrm{deg}}^{(\ell)}-A_{\mathrm{adj}}^{(\ell)}$, where $A_{\mathrm{adj}}^{(\ell)}$ is the filtered weighted adjacency matrix with a zero diagonal and $D_{\mathrm{deg}}^{(\ell)}$ is its degree matrix. Let $\{\lambda_k\}_{k=1}^{N}$ denote the eigenvalues of $\mathcal{L}^{(\ell)}$ in ascending order. Six quantities were computed at each level: 
(i) the harmonic multiplicity,
\begin{equation}
m_0
=
\left|\left\{
k:|\lambda_k|<\eta_{\lambda}\cdot\max(|\lambda_N|,1),
\ 1\leq k\leq N
\right\}\right|,
\end{equation}
which counts the zero eigenvalues within the relative spectral tolerance $\eta_{\lambda}=10^{-8}$ and equals the number of connected graph components at that filtration level; 
(ii) the smallest positive non-harmonic eigenvalue,
\begin{equation}
\lambda_{\min}^{+}
=
\lambda_{m_0+1}
=
\min\left\{
\lambda_k:\lambda_k\geq\eta_{\lambda}\cdot\max(\lambda_N,1)
\right\};
\end{equation}
(iii) the arithmetic mean of the non-zero eigenvalues,
\begin{equation}
\bar{\lambda}
=
\frac{1}{N-m_0}
\sum_{k=m_0+1}^{N}\lambda_k;
\end{equation}
(iv) the standard deviation of the non-zero eigenvalues,
\begin{equation}
\sigma_{\lambda}
=
\sqrt{
\frac{1}{N-m_0}
\sum_{k=m_0+1}^{N}
(\lambda_k-\bar{\lambda})^2
};
\end{equation}
(v) the maximum eigenvalue,
\begin{equation}
\lambda_{\max}=\lambda_N;
\end{equation}
and (vi) the sum of all eigenvalues, equal to the trace of the Laplacian,
\begin{equation}
\Sigma_{\lambda}
=
\sum_{k=1}^{N}\lambda_k
=
\operatorname{Tr}(L).
\end{equation}
These six descriptors summarize complementary properties of the graph Laplacian spectrum. The zero-eigenvalue multiplicity $m_0$ gives the number of connected components. The smallest positive eigenvalue $\lambda_{\min}^{+}$ characterizes the lowest nonzero spectral scale and equals the algebraic connectivity $\lambda_2$ when the graph is connected. The mean $\bar{\lambda}$ and standard deviation $\sigma_{\lambda}$ describe the location and dispersion of the nonzero spectrum, while $\lambda_{\max}$ specifies its upper bound. The eigenvalue sum $\Sigma_{\lambda}$ equals the Laplacian trace and the sum of weighted node degrees, or equivalently twice the total edge weight.

Per-atom descriptors were obtained through leave-one-out graph perturbations. Each C$\alpha$ atom $i$ was removed in turn, and the six spectral quantities were recomputed at every filtration level. Their absolute deviations from the full-graph values were then assigned to atom $i$. Concatenating the six perturbation sensitivities across the ten filtration levels produced a 60-dimensional spectral perturbation vector per atom. Combining this vector with the 60-dimensional dynamics block yielded the 120-dimensional LOPO representation.

Four regression models with fixed configurations were evaluated, including GBDT, RF, RR, and KNN. Each was trained on all proteins except the held-out test protein and evaluated on that protein. No grid search or feature standardization was applied. Five LOPO cycles with different random seeds were averaged. The full four-model comparison and the fold-specific per-level $\varepsilon$ schedules are reported in Tables~S7 and S5 of the Supporting Information, respectively.

\subsection{Single-cell representation and classification protocol}\label{subsec:scrna-methods}

Thirteen publicly available scRNA-seq datasets were used, each containing 300 to 1,937 cells with known cell-type labels. For each dataset, cells with fewer than 200 detected genes were filtered out, and genes detected in fewer than 3 cells were removed. The expression matrix was normalized to 10,000 counts per cell and log-transformed ($\ln(1+x)$). The top 2,000 HVGs were then selected using the Seurat flavor implemented in Scanpy, after which the resulting matrix was z-score standardized and projected into 30 principal components via PCA.

Multi-scale graphs were constructed from pairwise Euclidean distances between cells in the 30-dimensional PC space. Ten filtration levels were defined using equally spaced percentiles ranging from the 5th to the 95th percentile of the distance distribution. Let $d_{ij}^{(\ell)}$ denote the Euclidean distance between cells $i$ and $j$. At each level $\ell$, the off-diagonal adjacency weights were $(B_{\mathrm{adj}}^{(\ell)})_{ij}=\exp(-d_{ij}/\sigma)$ for retained cell pairs and zero otherwise, with $\sigma$ set to the median non-zero retained distance and $(B_{\mathrm{adj}}^{(\ell)})_{ii}=0$. As in the protein analysis, graph coupling used the standard Laplacian $\mathcal{L}^{(\ell)}=D_{\mathrm{deg}}^{(\ell)}-B_{\mathrm{adj}}^{(\ell)}$, or equivalently the zero-row-sum generator $-\mathcal{L}^{(\ell)}$.

For each coupling strength and filtration level, the impulse-response trajectory of each cell was recorded. Six cell-specific Laplacian spectral perturbation features were also computed. Each cell was removed in turn, and their absolute deviations from the corresponding full-graph values were then recorded. Concatenating the six dimensional statistical dynamical features and six dimensional spectral perturbation features across the ten filtration levels yielded a 120-dimensional representation. The 30 cell-specific PC scores, representing the coordinates of each cell along PC1--PC30, were appended to obtain the final 150-dimensional feature vector.

Cell-type classification used the ML4 procedure. For each $\varepsilon$, logistic regression (LogReg), support vector machine (SVM), RF, and KNN were evaluated separately. Features were standardized within each CV fold. During  fold-wise training-set adjustment, each eligible class was sampled with replacement to a common target sample size. The pooled sample indices were then deduplicated, yielding a training subset without repeated rows. Stratified five-fold CV with ten independent repeats was used, yielding 50 fold-level evaluations, with BA as the primary metric. For each dataset, the combination of coupling strength and classifier yielding the highest mean BA across the repeated CV folds was selected. The same repeated CV results were used for both configuration selection and performance summarization. No nested outer CV was performed. The class-eligibility rule, resampling target, and ML4 parameters are reported in Table~S17 of the Supporting Information. The architecture and training settings of the LSTM comparator using the 150-dimensional feature representation are reported in Table~S18 of the Supporting Information. 

We want to emphasize that the scRNA-seq experiment was conducted as a transductive within-dataset node-classification task. All cells participated in unsupervised PCA, graph construction, and response-feature generation, whereas cell-type labels were used only within the subsequent CV procedure.

\section{Conclusion}\label{sec:conclusion}

This study introduced a multiscale graph-coupled Burgers framework for constructing predictive node descriptors from protein structures and single-cell transcriptomic profiles. Filtration defines structural neighborhoods at different scales, within which controlled nonlinear dynamics generate node-specific impulse responses. Statistical dynamical features summarize these temporal response patterns in a compact form. For cross-protein prediction and single-cell classification, spectral perturbation features additionally characterize how node removal changes the graph spectrum, while PC scores preserve explicit expression-space coordinates in the single-cell representation. These features provide different descriptions of node environments within a shared computational framework.

The protein results showed that compact response statistics achieved within-protein correlations comparable to those obtained with densely sampled trajectories, suggesting that increasing temporal detail did not yield a clear advantage in mean PCC across the three datasets. Positive mean correlations under leave-one-protein-out evaluation further indicated that some predictive associations transferred across proteins, although performance varied among structures and depended on the dataset and comparator. In single-cell classification, the ablations supported combining information across filtration levels and retaining PC scores alongside graph-derived descriptors. These findings identify useful representation choices without establishing the independent predictive contribution of each dynamical or spectral component.

The principal contribution of our work is a strategy for translating static biological neighborhoods into predictive computational responses, with applications to both structural and transcriptomic data. The current evidence concerns configuration-selected within-protein prediction, cross-protein evaluation within the studied collections, and transductive cell classification. In the future, independent evaluation, targeted component ablations, and assessment of cell-order sensitivity would help establish the reliability and broader applicability of the present framework. This approach provides a basis for investigating how multiscale connectivity and controlled nonlinear responses can be combined to construct informative biological representations.

\section*{Supporting information}
Supporting information accompanies this manuscript and contains extended methodological details, complete parameter tables, and supplementary experimental results.

\section*{Data and code availability}
The source code and data files underlying this work are available at \url{https://github.com/Wzhgeek/BurgersBfactors}.

\section*{Acknowledgments}
This work was supported in part by Wuhan Textile University. The work of M.Z., X.W. and XXW was
supported in part by the MSU Research Foundation. The authors thank Dr. Guo-Wei Wei for discussion.  

\bibliographystyle{abbrv}
\bibliography{ref}

@article{wei2002conjugate,
  title={Conjugate filter approach for solving {Burgers’} equation},
  author={Wei, GW and Gu, Yun},
  journal={Journal of Computational and Applied mathematics},
  volume={149},
  number={2},
  pages={439--456},
  year={2002},
  publisher={Elsevier}
}

@article{cang2020eh,
  author  = {Cang, Zixuan and Munch, Elizabeth and Wei, Guo-Wei},
  year    = {2020},
  month   = {12},
  title   = {Evolutionary Homology on Coupled Dynamical Systems with Applications to Protein Flexibility Analysis},
  journal = {Journal of Applied and Computational Topology},
  doi     = {10.1007/s41468-020-00057-9}
}

@article{xia2014fri,
  author  = {Opron, Kristopher and Xia, Kelin and Wei, Guo-Wei},
  year    = {2014},
  month   = {06},
  pages   = {234105},
  title   = {Fast and Anisotropic Flexibility-Rigidity Index for Protein Flexibility and Fluctuation Analysis},
  volume  = {140},
  journal = {The Journal of Chemical Physics},
  doi     = {10.1063/1.4882258}
}

@article{xia2014mnd,
  author  = {Xia, Kelin and Wei, Guo-Wei},
  year    = {2014},
  month   = {03},
  pages   = {013103},
  title   = {Molecular Nonlinear Dynamics and Protein Uncertainty Quantification},
  volume  = {24},
  journal = {Chaos},
  doi     = {10.1063/1.4861202}
}

@article{park2013gnm,
  author  = {Park, Jun-Koo and Jernigan, Robert L. and Wu, Zhi-jun},
  title   = {Coarse Grained Normal Mode Analysis vs. Refined Gaussian Network Model for Protein Residue-Level Structural Fluctuations},
  journal = {Bulletin of Mathematical Biology},
  year    = {2013},
  volume  = {75},
  pages   = {124--160}
}

@article{wang2020psg,
  author  = {Wang, Rui and Nguyen, Duc Duy and Wei, Guo-Wei},
  title   = {Persistent Spectral Graph},
  journal = {International Journal for Numerical Methods in Biomedical Engineering},
  year    = {2020},
  volume  = {36},
  number  = {9},
  doi     = {10.1002/cnm.3376}
}

@article{hayes2025psl,
  author  = {Hayes, Nicole and Wei, Xiaoqi and Feng, Hongsong and Merkurjev, Ekaterina and Wei, Guo-Wei},
  title   = {Persistent Sheaf Laplacian Analysis of Protein Flexibility},
  journal = {The Journal of Physical Chemistry B},
  year    = {2025},
  volume  = {129},
  number  = {17},
  pages   = {4169--4178},
  doi     = {10.1021/acs.jpcb.5c01287}
}

@article{sun2019bfactor,
  author  = {Sun, Zhoutong and Liu, Qian and Qu, Ge and Feng, Yan and Reetz, Manfred T.},
  title   = {Utility of {B}-Factors in Protein Science: Interpreting Rigidity, Flexibility, and Internal Motion and Engineering Thermostability},
  journal = {Chemical Reviews},
  year    = {2019},
  volume  = {119},
  number  = {3},
  pages   = {1626--1665},
  doi     = {10.1021/acs.chemrev.8b00290}
}

@article{pandey2023b,
  title={B-factor prediction in proteins using a sequence-based deep learning model},
  author={Pandey, Akash and Liu, Elaine and Graham, Jacob and Chen, Wei and Keten, Sinan},
  journal={Patterns},
  volume={4},
  number={9},
  year={2023},
  publisher={Elsevier}
}

@article{bramer2018blind,
  title={Blind prediction of protein B-factor and flexibility},
  author={Bramer, David and Wei, Guo-Wei},
  journal={The Journal of chemical physics},
  volume={149},
  number={13},
  year={2018},
  publisher={AIP Publishing}
}

@article{tan2019singlecellnet,
  title={SingleCellNet: a computational tool to classify single cell RNA-Seq data across platforms and across species},
  author={Tan, Yuqi and Cahan, Patrick},
  journal={Cell systems},
  volume={9},
  number={2},
  pages={207--213},
  year={2019},
  publisher={Elsevier}
}

@article{wang2024scfed,
  title={scFed: federated learning for cell type classification with scRNA-seq},
  author={Wang, Shuang and Shen, Bochen and Guo, Lanting and Shang, Mengqi and Liu, Jinze and Sun, Qi and Shen, Bairong},
  journal={Briefings in Bioinformatics},
  volume={25},
  number={1},
  pages={bbad507},
  year={2024},
  publisher={Oxford University Press}
}

@article{jiang2025,
  title = {Machine Learning Predictions from Unpredictable Chaos},
  author = {Jiang, Jian and Chen, Long and Ke, Lu and Dou, Bozheng and Zhu, Yueying and Shi, Yazhou and Qiu, Huahai and Zhang, Ben-gong and Zhou, Tianshou and Wei, Guo-Wei},
  year = 2025,
  journal = {Journal of The Royal Society Interface},
  volume = {22},
  number = {231},
  eprint = {2503.14956},
  pages = {20250441},
  issn = {1742-5662},
  doi = {10.1098/rsif.2025.0441},
  archiveprefix = {arXiv},
  pmid = {41027482}
}

@article{barthels2021bandeltait,
  title={BAN$\Delta$IT: B’-factor analysis for drug design and structural biology},
  author={Barthels, Fabian and Schirmeister, Tanja and Kersten, Christian},
  journal={Molecular informatics},
  volume={40},
  number={1},
  pages={2000144},
  year={2021},
  publisher={Wiley Online Library}
}

@article{erman2025extending,
  title={Extending the Gaussian network model: integrating local, allosteric, and structural factors for improved residue-residue correlation analysis},
  author={Erman, Burak},
  journal={Physical Biology},
  volume={22},
  number={6},
  pages={066010},
  year={2025},
  publisher={IOP Publishing}
}

@article{xia2015multiscale,
  title={Multiscale Gaussian network model (mGNM) and multiscale anisotropic network model (mANM)},
  author={Xia, Kelin and Opron, Kristopher and Wei, Guo-Wei},
  journal={The Journal of chemical physics},
  volume={143},
  number={20},
  year={2015},
  publisher={AIP Publishing}
}

@article{zheng2008unification,
  title={A unification of the elastic network model and the Gaussian network model for optimal description of protein conformational motions and fluctuations},
  author={Zheng, Wenjun},
  journal={Biophysical Journal},
  volume={94},
  number={10},
  pages={3853--3857},
  year={2008},
  publisher={Elsevier}
}

@article{nguyen2016generalized,
  title={Generalized flexibility-rigidity index},
  author={Nguyen, Duc Duy and Xia, Kelin and Wei, Guo-Wei},
  journal={The Journal of chemical physics},
  volume={144},
  number={23},
  year={2016},
  publisher={AIP Publishing}
}

@article{hayes2026predicting,
  title={Predicting protein--nucleic acid flexibility using persistent sheaf Laplacians},
  author={Hayes, Nicole and Merkurjev, Ekaterina and Wei, Guo-Wei},
  journal={Physical Chemistry Chemical Physics},
  volume={28},
  number={3},
  pages={2215--2226},
  year={2026},
  publisher={The Royal Society of Chemistry}
}

@inproceedings{cheng20254d,
  title={4D diffusion for dynamic protein structure prediction with reference and motion guidance},
  author={Cheng, Kaihui and Liu, Ce and Su, Qingkun and Wang, Jun and Zhang, Liwei and Tang, Yining and Yao, Yao and Zhu, Siyu and Qi, Yuan},
  booktitle={Proceedings of the AAAI Conference on Artificial Intelligence},
  volume={39},
  number={1},
  pages={93--101},
  year={2025}
}

@article{zhang2025flexible,
  title={Flexible protein--ligand docking with diffusion-based side-chain packing},
  author={Zhang, Runze and Jiang, Xinyu and Cao, Duanhua and Wang, Zhaokun and Yu, Jie and Chen, Mingan and Fan, Zhehuan and Kong, Xiangtai and Xiong, Jiacheng and Zhang, Zimei and others},
  journal={Proceedings of the National Academy of Sciences},
  volume={122},
  number={52},
  pages={e2511925122},
  year={2025},
  publisher={National Academy of Sciences}
}

@article{burgers1948mathematical,
  title={A mathematical model illustrating the theory of turbulence},
  author={Burgers, Johannes Martinus},
  journal={Advances in applied mechanics},
  volume={1},
  pages={171--199},
  year={1948},
  publisher={Elsevier}
}

@article{zeng2025cellfm,
  title={CellFM: a large-scale foundation model pre-trained on transcriptomics of 100 million human cells},
  author={Zeng, Yuansong and Xie, Jiancong and Shangguan, Ningyuan and Wei, Zhuoyi and Li, Wenbing and Su, Yun and Yang, Shuangyu and Zhang, Chengyang and Zhang, Jinbo and Fang, Nan and others},
  journal={Nature Communications},
  volume={16},
  number={1},
  pages={4679},
  year={2025},
  publisher={Nature Publishing Group UK London}
}

@article{du2026scalgsl,
  title={scALGSL: Active Learning and Graph Structure Learning for Cell Type Annotation From Single-Cell RNA-seq Data},
  author={Du, Zhi-Hua and Yi, Jia-Le and Hu, Wei-Lin and Li, Jian-Qiang and You, Hai-Ru and You, Zhu-Hong and Huang, Zhi-An and Huang, Yu-An},
  journal={IEEE Transactions on Computational Biology and Bioinformatics},
  year={2026},
  publisher={IEEE}
}

@article{long2026scannet,
  title={ScanNet: Single-cell annotation informed by transcriptional regulation Network via iterative heterogeneous graph learning},
  author={Long, Yongyu and Zhang, Wenhao and Cao, Lan and Huang, Xiaobing and Wang, Ying},
  journal={PLOS Computational Biology},
  volume={22},
  number={8},
  pages={e1014602},
  year={2026},
  publisher={Public Library of Science San Francisco, CA USA}
}

@article{feng2025mdg,
  author  = {Feng, Hongsong and Zhao, Jeffrey Y. and Wei, Guo-Wei},
  title   = {Multiscale Differential Geometry Learning for Protein Flexibility Analysis},
  journal = {Journal of Computational Chemistry},
  year    = {2025},
  volume  = {46},
  number  = {7},
  pages   = {e70073},
  doi     = {10.1002/jcc.70073}
}

@article{bi2025topomag,
  author  = {Bi, Wanying and Feng, Hongsong and Wu, Jie and Li, Jingyan and Wei, Guo-Wei},
  title   = {Topological Magnitude for Protein Flexibility Analysis},
  journal = {Royal Society Open Science},
  year    = {2025},
  volume  = {12},
  number  = {12},
  pages   = {251599},
  doi     = {10.1098/rsos.251599}
}

@article{suTopo2025,
  title = {Topological Data Analysis and Topological Deep Learning beyond Persistent Homology: {{A}} Review},
  author = {Su, Zhe and Liu, Xiang and Hamdan, Layal Bou and Maroulas, Vasileios and Wu, Jie and Carlsson, Gunnar and Wei, Guo-Wei},
  year = 2025,
  journal = {Artificial Intelligence Review},
  volume = {59},
  number = {2},
  pages = {58},
  issn = {1573-7462},
  doi = {10.1007/s10462-025-11462-w},
  urldate = {2026-01-16},
  langid = {english}
}

@article{weeRev2025,
  title = {A {{Review}} of {{Topological Data Analysis}} and {{Topological Deep Learning}} in {{Molecular Sciences}}},
  author = {Wee, Junjie and Jiang, Jian},
  year = 2025,
  journal = {J. Chem. Inf. Model},
  volume = {65},
  pages = {12691--12706},
  doi = {10.1021/acs.jcim.5c02266}
}

@book{Burgers1974,
  author    = {Burgers, J. M.},
  title     = {The Nonlinear Diffusion Equation},
  year      = {1974},
  publisher = {Springer Netherlands},
  address   = {Dordrecht},
  doi       = {10.1007/978-94-010-1745-9}
}

@article{xiaPersist2014,
  title = {Persistent Homology Analysis of Protein Structure, Flexibility, and Folding},
  author = {Xia, Kelin and Wei, Guo-Wei},
  year = 2014,
  journal = {International Journal for Numerical Methods in Biomedical Engineering},
  volume = {30},
  number = {8},
  pages = {814--844},
  issn = {2040-7939, 2040-7947},
  doi = {10.1002/cnm.2655},
  urldate = {2026-05-06},
  copyright = {http://onlinelibrary.wiley.com/termsAndConditions\#vor},
  langid = {english}
}

@article{xiaMulti2015,
  title = {Multiresolution Persistent Homology for Excessively Large Biomolecular Datasets},
  author = {Xia, Kelin and Zhao, Zhixiong and Wei, Guo Wei},
  year = 2015,
  journal = {Journal of Chemical Physics},
  volume = {143},
  number = {13},
  issn = {10897690},
  doi = {10.1063/1.4931733},
  pmid = {26450288}
}

@article{wei2025persistent,
  title={Persistent topological laplacians—a survey},
  author={Wei, Xiaoqi and Wei, Guo-Wei},
  journal={Mathematics},
  volume={13},
  number={2},
  pages={208},
  year={2025},
  publisher={MDPI}
}

\end{document}